\documentclass[preprint,floats,aps,12pt,superscriptaddress,nofootinbib,floatfix]{revtex4}
\usepackage{natbib} 
\usepackage{times} 
\usepackage{amssymb,amsmath}
\usepackage{csquotes}

\usepackage{algorithm}
\usepackage{algpseudocode}
\usepackage[usenames,dvipsnames]{color}
\usepackage{epstopdf}
\usepackage[pdftex]{graphicx}
\usepackage[pdftex]{hyperref}
\hypersetup{a4paper,
    pdftitle={Manuscript on XXX} 
    pdfauthor={Tim W. Kroll, Oliver Kamps},  
pdfproducer={lateX},
pdfview=FitV,       
pdfstartview=FitB,
linkcolor=blue,     
citecolor=blue,     
urlcolor=red,      
breaklinks=true,    
colorlinks=true,
citebordercolor=0 0 0,  
filebordercolor=0 0 0,
linkbordercolor=0 0 0,
menubordercolor=0 0 0,
urlbordercolor=0 0 0,
pdfhighlight=/I,
pdfborder=0 0 0,   
bookmarksopen=true,
bookmarksnumbered=true
}
\DeclareGraphicsExtensions{.jpg, .pdf, .tif, .png}
\begin{document}
%
\title{Sparse identification of evolution equations via bayesian model selection}
\author{Tim W. Kroll}
\email{tim.kroll@uni-muenster.de}
\affiliation{Institute of Theoretical Physics, University of M\"unster, Wilhelm-Klemm-Str.\ 9, 48149 M\"unster, Germany}
\affiliation{Center for Nonlinear Science (CeNoS), University of M\"unster, Corrensstr.\ 2, 48149 M\"unster, Germany}
\author{Oliver Kamps}
\affiliation{Center for Nonlinear Science (CeNoS), University of M\"unster, Corrensstr.\ 2, 48149 M\"unster, Germany}
\begin{abstract}
The quantitative formulation of evolution equations is the backbone for prediction, control, and understanding of dynamical systems across diverse scientific fields.  Besides deriving differential equations for dynamical systems based on basic scientific reasoning or prior knowledge in recent times a growing interest emerged to infer these equations purely from data. In this article, we introduce a novel method for the sparse identification of nonlinear dynamical systems from observational data, based on the observation how the key challenges of the quality of time derivatives and sampling rates influence this general problem. Our approach combines system identification based on thresholded least squares minimization with additional error measures that account for both the deviation between the model and the time derivative of the data, and the integrated performance of the model in forecasting dynamics. Specifically, we integrate a least squares error as well as the Wasserstein metric for estimated models and combine them within a Bayesian optimization framework to efficiently determine optimal hyperparameters for thresholding and weighting of the different error norms. Additionally, we employ distinct regularization parameters for each differential equation in the system, enhancing the method's precision and flexibility.

We demonstrate the capabilities of our approach through applications to dynamical fMRI data and the prototypical example of a wake flow behind a cylinder. In the wake flow problem, our method identifies a sparse, accurate model that correctly captures transient dynamics, oscillation periods, and phase information, outperforming existing methods. In the fMRI example, we show how our approach extracts insights from a trained recurrent neural network, offering a novel avenue for explainable AI by inferring differential equations that capture potentially causal relationships. 
\end{abstract}
\maketitle
%
%
%
\section{Introduction} \label{sec:intro}
The formulation of evolution equations for dynamical systems in the form of coupled (non)linear differential equations is a powerful scientific tool that has been adapted successfully in different fields of science \cite{strogatz2014nonlinear} far beyond its origins in classical mechanics \cite{newton1687philosophiae}. Examples include the mathematical modeling of climate dynamics \cite{Lucarini2020}, power grids \cite{Witthaut2022}, or medicine \cite{Davis2019}, to name just a few.

The rapid advancements in machine learning and the growing availability of data over the past decade have led the scientific community to increasingly integrate data-driven approaches into the scientific process \cite{Wang2023}. These developments have sparked significant interest in leveraging such techniques, along with other data-driven methods, for the analysis and modeling of dynamical systems. For instance, Long Short-Term Memory (LSTM) networks have been successfully applied to forecast complex systems \cite{Vlachas2018, Wan2018pone}.

The recent rise of Transformer models presents a natural extension of these methods; however, their applicability to time-series forecasting remains a topic of active debate \cite{Zeng2023}. Furthermore, approaches grounded in dynamical systems, such as cluster modeling \cite{Fernex2021} and Koopman theory \cite{Mezic2021}, have also gained substantial attention in recent years. A comprehensive overview of these techniques is provided in \cite{CampsValls2023}. Additionally, modified recurrent neural networks offer another promising direction, as demonstrated in a recent study \cite{Durstewitz2023nature}.

Nevertheless, most of these methods can be regarded as black-box approaches \cite{Succi2019pta}, which are often limited in their ability to identify causal structures as opposed to mere correlations \cite{Succi2019pta}. This limitation poses a significant challenge when applying such methods to many scientific questions. Consequently, it appears advantageous to combine the availability of data and advancements in machine learning techniques with the formulation of scientific relations through differential equations. This approach, often characterized as symbolic artificial intelligence \cite{Jimenez2018jfm,Jimenez2020arxiv}, offers the benefit of uncovering causal structures inherent to the scientific problem. 

Regarding dynamical systems, this field has made significant progress in recent years, exemplified by the application of genetic algorithms \cite{Bongard2007, Schmidt2009}. A notable milestone was the introduction of a computationally efficient method for the sparse identification of dynamical systems \cite{Brunton2016pnas} called SINDy, along with its extensions for ordinary differential equations (ODEs), such as the inclusion of constraints \cite{Loiseau2018} and implicit models for biological systems \cite{Mangan2016}. The primary advantage of the resulting sparse models is their enhanced interpretability. As a result, these methods have been successfully applied across a diverse range of fields, including fluid dynamics \cite{loiseau:hal-02398729}, reaction dynamics in chemistry \cite{Hoffmann2019}, epidemiology \cite{Horrocks2020}, plasma physics \cite{Dam2017}, and protein networks in bioinformatics \cite{Pantazis2018}, among others. For a comprehensive overview of data-driven methods for dynamical systems, see \cite{North2023}.

Nevertheless, when applying an algorithm like SINDy \cite{Brunton2016pnas} to real-world data or in complex settings, several challenges may arise. For instance, in \cite{Dam2017}, the algorithm was not able to find a model for a limit cycle in an application from plasma physics, although this seems to be a rather straightforward problem setting.  Another challenge is determining the appropriate level of sparsity for the final model. In fluid dynamical models studied in \cite{supekar2023pnas}, this issue is addressed by comparing results with analytical solutions. However, such comparisons may not always be feasible, particularly for systems lacking an established analytical benchmark. To make use of SINDy as part of an algorithm to estimate covariant Lyapunov vectors \cite{Martin2022}, a very fine time sampling was necessary. This constraint reduces the method’s practicability for real-world data, where experimental limitations may prevent achieving such time resolution. Even in numerical studies, small time steps might be impractical due to the trade-off between computational accuracy and feasibility.

To overcome the aforementioned problems and thereby to widen the applicability of system identification methods, we develop a robust but computationally very efficient algorithm for reconstructing sparse models from data that relies on the combination of different error measures.   

\begin{figure*}[t]
\centering
\includegraphics[width=\linewidth]{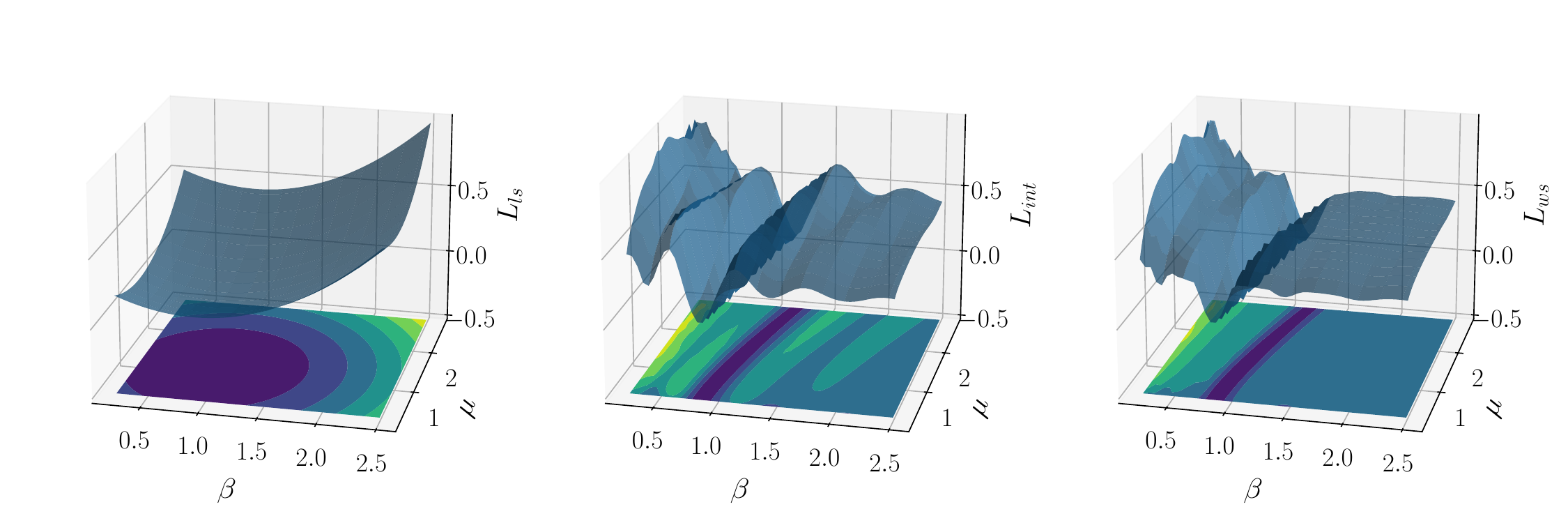}
\caption{The potential plots for different error functions. We calculated the errors for an interval of $\pm 0.05$ around the correct values of $\mu=1.0$ and $\beta=1.0$ for the van der Pol Oscillator. The least squares error was calculated for exact derivatives. (left) The convex error function of eq. \ref{eq:xdoterror} shows only one minimum, (middle) the non-convex error function of eq. \ref{eq:xerror} show different minima of the same problem, where the global minimum corresponds to the values of the original equations, (right) the non-convex error function of eq. \ref{eq:W2} exhibits a global minimum, which again corresponds to the correct values of the coefficients, but has fewer local minima
\label{fig:potentiale_vergleich} }
\end{figure*}

\section{Identification of model dynamics}

Starting point is the hypothesis that an observed multidimensional time series $\mathbf{y}$ can be approximated by the solution of a set of coupled nonlinear differential equations
\begin{equation}
\frac{d}{dt}\mathbf{x}\left(t\right)=\mathbf{f}\left(\mathbf{x},\mathbf{\sigma},t\right),
\label{eq:state}
\end{equation}
where $\mathbf{x}\in\mathbb{R}^n$ is the state vector of a system and $\mathbf{f}\left(\mathbf{x},\mathbf{\sigma},t\right)$ represents the dynamical law that is depending on the state $\mathbf{x}$ itself, possibly on the time $t$ and is parameterized by a set  of parameters $\mathbf{\sigma}$. From now on, we will drop the dependence on $t$ as the following ideas can easily be generalized to nonautonomous systems. Our goal is to estimate an analytical description of $\mathbf{f}$ from data. To this end, we have to introduce an ansatz for the right hand side which has to be as general as needed to sufficiently capture the dynamics of the system. In analogy to systems of nonlinear differential equations used in many fields of science we will choose a linear combination of some analytical functions $\xi_0,\xi_1,...\xi_N$ and define the components $f_i$ of $\mathbf{f}\left(\mathbf{x},\mathbf{\sigma}\right)$ as:
\begin{equation}
    f_i=\sum_j \sigma_{ij}\xi_j.
    \label{eq:linearAnsatz}
\end{equation}

Possible right hand side functions are for example polynomials of different order, trigonometric functions or combinations of those and will be defined separately for every explicit application. In general, the underlying equations are assumed to depend only on a few terms, since very simple equations can lead to a plethora of different behaviours, as we can see, i.e. in the Lorenz system \cite{Lorenz63jas}. \\

To illustrate and guide the thought process we use in the following the example of the nonlinear van der Pol Oscillator \cite{vanderPol1926}, which is defined by the following differential equations
\begin{align}
\label{eq:VdP}
    \dot{x_0}&=\beta x_1 \\
    \dot{x_1}&=\mu\left(1-x_0^2\right)x_1-x_0\nonumber .
\end{align}
By using a suitable ansatz, the identification of the right hand side $\mathbf{f}$ in \eqref{eq:state} or for the concrete example \eqref{eq:VdP} is  boiled down to the problem of finding the coefficients $\sigma_{ij}$. This problem can be solved by introducing an objective function and finding its minimum corresponding to the best parameters. 

\begin{figure*}[ht]
\includegraphics[width=\linewidth]{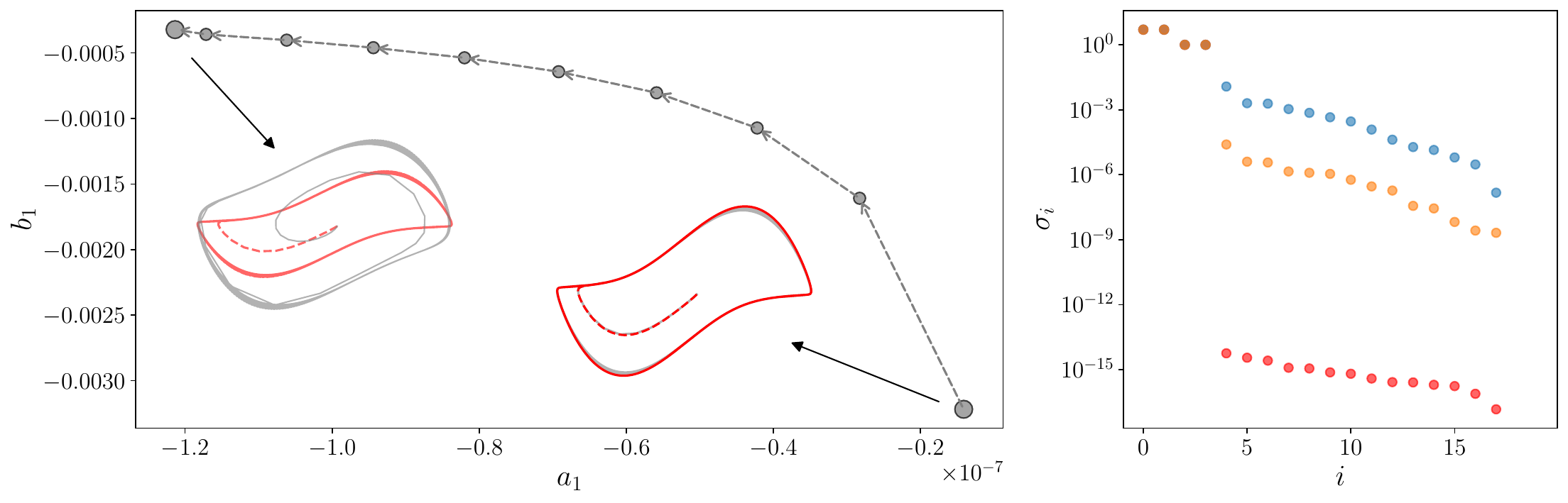}
\caption{(left) Comparison of the estimated values of spurious terms for the simplified van der Pol problem of equation \ref{eq:simpleVDP}. With lower sampling rates the spurious terms change values and the corresponding estimated attractors are shown for the two extreme sampling rates. The resemblance of the estimated attractor changes with the lowering of the sampling rate and dramatically deviates from the original nonlinear attractor of the van der Pol oscillator. (right) Change of absolute values of the full van der Pol problem for different methods of derivative approximation. While there is a distinct difference between the correct values of the coefficients for the exact derivatives, this difference gets smaller and smaller with the accuracy of the chosen methods. Spline derivatives still show some difference, but for gradient derivatives this gets smaller. \label{fig:spurious} }
\end{figure*}

One possibility is to minimize the difference between the left and right hand side of equation \eqref{eq:state} or \eqref{eq:VdP}. The error function can then be defined as:
\begin{equation}
    L_\text{ls} = \left\langle\sum_i\left(\dot{x}_i\left(t\right)-\sum_j\sigma_{ij}\xi_j\right)^2\right\rangle,
    \label{eq:xdoterror}
\end{equation}
The big advantage of this formulation is that the so defined error function is convex  (see left part of fig \ref{fig:potentiale_vergleich}). Especially in the case of a linear dependence of the ansatz on the parameters, the minimum of this function can be calculated analytically. 
While this approach is very appealing in theory, it can lead to problems in practical applications, as the quality of the result depends on the accuracy of $\dot{x}_i$. Since this quantity must be calculated numerically from the time series, it may be affected by the quality of the differentiation scheme or the sampling of the time series. To illustrate the consequences we try to estimate \eqref{eq:VdP} from data using the very restricted ansatz 
\begin{align}
    \dot{x_0}&=a_1 x_1 + a_2 x_1^2 \\
    \dot{x_1}-x_0&=b_1 \left(1-x_0^2\right)x_1 + b_2 x_1^2
    \label{eq:simpleVDP}
\end{align}
where only $a_1,a_2,b_1,b_2$ have to be estimated. We integrate \eqref{eq:VdP} for $\mu=5$, sample at various time scales, and estimate parameters using least squares based on \eqref{eq:xdoterror}. The plot in the left part of fig. \ref{fig:spurious} shows that as the sampling intervals increase, the numerical values of the terms $a_2$ and $b_2$, which should be zero, also grow, leading to a structural distortion of the attractor. If we further lower the sampling rate, we even get a change in the sign of the $b_2$-value (not shown). 
This means that although the estimated coefficients correspond to the minimum of \eqref{eq:xdoterror}, integrating this model may lead to a wrong attractor. Consequently, it reinforces the impression that \eqref{eq:xdoterror} is not suitable for estimating models from real data.

Another error source, which strongly interacts with the previous one, is the scheme for the numerical derivative. To demonstrate this, we estimate the van der Pol equation from a larger library of potential right-hand side terms using polynomials in $x_0$ and $x_1$ up to order three. We calculate derivatives using three methods: the exact derivative (from the differential equation), finite differences (first-order gradient), and splines (approximating the time series). Again, the coefficients are estimated through a simple least squares fit. The right part of figure \ref{fig:spurious} shows how the estimated coefficients differ across methods. The exact derivatives provide clear scale separation, allowing us to easily discard terms. However, in practice, we must rely on numerical derivatives. Spline-based derivatives still offer a workable scale separation, but finite differences show no clear separation, making it hard to set a threshold for discarding terms. But in case of even small noise the spline will also suffer from the same effect. These problems cannot always be avoided, as experiments are often constrained by the physical properties of the setup, limiting how finely we can sample. Even in numerical simulations, time and cost constraints may limit sampling resolution, requiring a balance between finer data and practical limitations.

In summary, while the error function has a correct minimum with exact derivatives, this minimum shifts when using approximated derivatives or sparser sampling, increasing spurious terms that are not part of the true differential equation and narrowing the gap between relevant and irrelevant terms. These spurious terms can hinder the interpretability of the estimated equations and cause significant distortions of the attractor, even when they are small.\\
One solution to neglect these terms is to regularize the error function with an L1-norm to ensure the number of terms to be small:
\begin{equation}
    L_{\lambda,ls}=L_\text{ls}+\lambda_{\text{L1}} \sum_{i,j}|\sigma_{ij}|.
        \label{eq:xdoterrorRegularized}
\end{equation}
In \cite{Brunton2016} the problem was effectively solved through sequential least squares thresholding. There, the least squares fit from equation \ref{eq:xdoterror} is solved without using regularization multiple times and every turn all terms below a threshold $\lambda$ are discarded and not used further.
The remaining key challenge is to select the appropriate threshold to balance between retaining relevant terms and removing spurious ones, even when there is no clear scale separation, while utilizing the numerical advantages of the SINDy method from \cite{Brunton2016}. For example in \cite{Mangan2017} the Akaike Information Criterion (AIC) based on the integrated timeseries from thresholded models was used to select an optimal threshold. But in the presented synthetic examples the time derivatives have been calculated via the known right hand side of the original models which leads to a pronounced scale separation in the coefficients, in principle making the usage of the AIC unnecessary in this case. The basic result from \cite{Mangan2017} could be summarized as showing that the AIC gets worse if we discard terms that are present in the original model. 

A straightforward approach to overcome the aforementioned problems is to use the L2-norm between the original time series and the integrated time series from the estimated model which is defined as:
\begin{equation}   L_\text{int}=\left\langle\left(\mathbf{x}\left(t\right)-\mathbf{x}\left(t_{1}\right)-\int_{t_{1}}^{t_{N_t}} \mathbf{f}(\mathbf{x}(\tau), \sigma) d \tau\right)^2\right\rangle
\label{eq:xerror}
\end{equation} 
The advantage of $L_\text{int}$ is that its minimum is exact and independent of sampling or derivative approximation. It directly shows whether small changes in the coefficients cause the time series to diverge. By focusing on the integrated time series, we can set a threshold to discard terms while still recovering the original data, eliminating spurious terms that would otherwise distort the model's trajectory. However, this comes with the challenge of a non-convex error landscape (see figure \ref{fig:potentiale_vergleich}), which is difficult to minimize. One method to address this is the multiple shooting method from \cite{Peifer2007ietsb}, which requires significantly more computational effort and does not a priori promote sparsity.

An alternative norm, working with the integrated time series is  the  Wasserstein-distance $L_{W2}$ defined as \cite{Kantorovich1960,Wasserstein1969,Cuturi2013}:
\begin{equation}
    L_{W2} = \inf _\pi\left(\frac{1}{n} \sum_{i=1}^n\left\|X_i-Y_{\pi(i)}\right\|^2\right)^{1 / 2} .
    \label{eq:W2}
\end{equation}
It is also sometimes denoted as Earth mover's distance and in a sense describes how costly it is to transport one trajectory into another in phase space. To see the advantages we analyze a two-dimensional dynamical system with a limit cycle, represented by two phase-shifted time series. When comparing these time series using $L_{int}$, we find a finite difference, suggesting they come from different systems, even though their underlying dynamics are identical. In contrast, the Wasserstein norm yields a difference of zero, accurately reflecting their shared dynamics. An additional advantage of $L_{W2}$ is its ability to avoid local minima of $L_{int}$ (see fig. \ref{fig:potentiale_vergleich}), which may correspond to constant values or fixed-point dynamics. These local minima are avoided because moving all mass from a fixed point to an oscillating trajectory would incur a significant cost.

\begin{figure*}[ht]
\centering
\includegraphics[width=\linewidth]{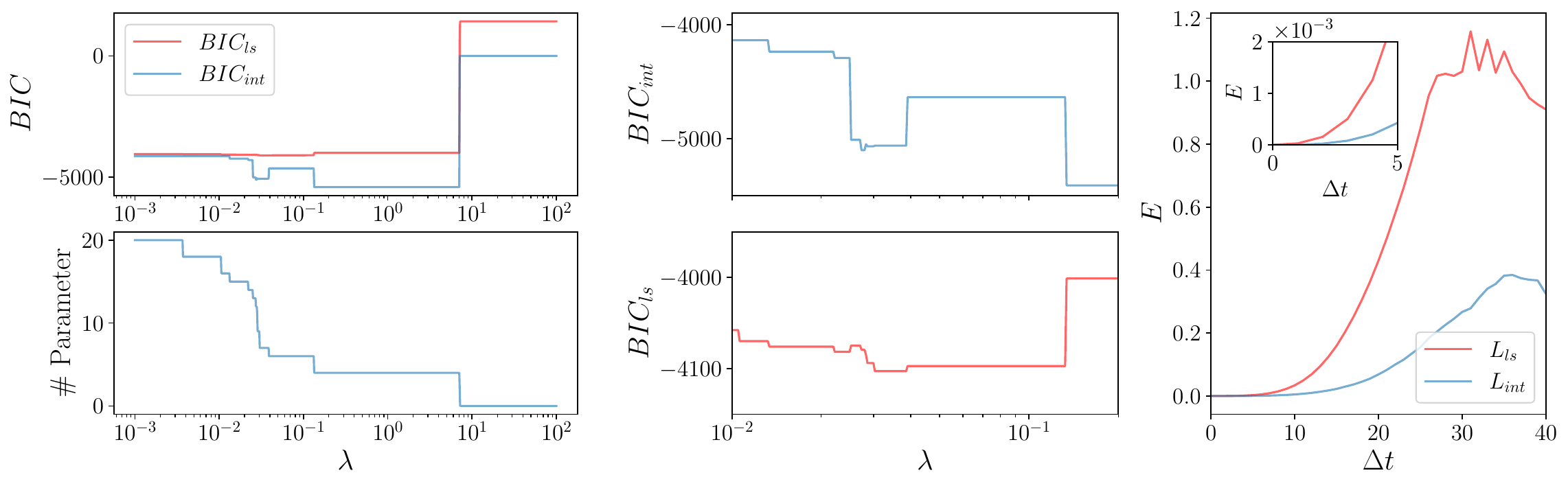}
\caption{(left,middle) Comparison of BIC and errors for SINDy and integrated error with respect to the number of parameters in the resulting models. The BIC is smaller for the integration error for all values of the threshold parameter. Furthermore the change in the BIC seems to occur for every change in parameter values, while an estimation based on the least squares error exhibits wide ranges of plateaus over several values of the number of coefficients.(right) Comparison of performance for the integration of estimated models for van der Pol with SINDy and rebel for different sampling rates. Even for small sampling rate the REBEL method outperforms the SINDy method. For higher sampling rates the accuracy deteriorates for both methods, while for REBEL the error is significantly lower. }
\label{fig:IC_error_vergleich}
\end{figure*}

\section{Reconstruction Error Based Estimation of dynamical Laws (REBEL)}

In summary in the proceeding section we have shown that the different error functions have complementary advantages and disadvantages. The remaining  question is how to combine their advantages to overcome the corresponding problems while keeping the method computational highly efficient. 

Relying on the convexity of \eqref{eq:xdoterror} we first perform a linear regression to find the model parameters $\sigma_{ij}$. Similar to \cite{Brunton2016} we discard all terms below a certain threshold $\lambda$. Since many nonlinear systems have different time scales, using just one $\lambda$ can lead to false results. As an example we can imagine a two-dimensional system, where we have different orders of magnitude for the coefficients. If we have coefficients of order 1 and of order $10^{-3}$, with spurious coefficients between those two orders in the first equation, we are stuck with the problem of either discarding all coefficients of the second one or allowing those spurious coefficients in the first one. It is easily seen, that in this way we are never able to recover the original equations. Therefore instead of thresholding all right hand side functions of an n-dimensional system with the same threshold $\lambda$, we use a different $\lambda_i$ for each function. 
The model found this way is now integrated once and compared to the original time series $x(t)$ via \eqref{eq:xerror} and \eqref{eq:W2}. To benefit from the advantages of both norms we combine them in analogy to the elastic net regularization \cite{Zou2005} in a mixed error function
\begin{equation}
L_{\text{mixed}} = \alpha L_{\text{int}} + (1-\alpha) L_{\text{W2}}
\end{equation}
where their influence is balanced by  the hyperparameter $\alpha$. In principle we now could follow the approach in \cite{Mangan2017} and construct a subset of sparse models  within a potentially vast combinatorial model space by performing the regularized least squares fit for certain choices of $\lambda$.  Furthermore, rather than limiting the search to a predefined subset, we explore a broader space of possible models, optimizing BIC to identify the best-fitting model. Using an information criterion like AIC used in \cite{Mangan2017} one can choose the model best balancing the quality of the fit and the sparsity. Here we use the Bayesian Information Criterion $BIC$ \cite{Schwarz1978} which is defined as
\begin{equation}
    BIC=k\ln(n)-2\ln(L)
    \label{eq:BIC}
\end{equation}
where $k$ is the number of free model parameters, here corresponding to the nonzero $\sigma_{ij}$, $n$ is the number of data points and $L$ is a likelihood (error) function. This criterion is derived and explained in the textbook \cite{Murphy2012} and can be used to do a Bayesian model selection balancing the accuracy and the complexity of a model.
Nevertheless, the approach to estimate  models based on different choices of $\lambda$ and afterwards choosing the best model based on an information criterion is not feasible anymore since we introduced the additional hyper parameter $\alpha$ and we want to use different $\lambda$'s for every equation in \eqref{eq:linearAnsatz}. In this case we would have to sample an $(n+1)$-dimensional model space.  Therefore, in our approach we choose an initial set of $\lambda$'s and an initial value for $\alpha$ to solve the identification problem using \eqref{eq:xdoterror} combined with thresholding. This leads to a sparse set of non-zero parameters $\sigma_{ij}$ in \eqref{eq:linearAnsatz} and a corresponding $k$. We now integrate the system once and compute $L_{\text{mixed}}$ using the initial $\alpha$. Now we can use an optimization algorithm to find the minimum of \eqref{eq:BIC} leading to a model where sparsity and adaption to the data are balanced in an optimal way, without sampling a high dimensional model space.

To get an impression of the behavior of the BIC we again consider  
the van der Pol system \eqref{eq:VdP} and make an ansatz for the right hand side including terms up to third order leading to 30 possible coefficients. The derivatives are computed via finite differences.  For simplicity, here we use only one $\lambda$ for both equations. We now estimate the parameters via sequential thresholding like in \cite{Brunton2016pnas} based on $L_{ls}$ leading to a number $k$ of non-zero coefficients. In one case we plug $L_{ls}$ into the BIC and in the second case we integrate the identified ODE system once, compute $L_{int}$ and plug this into the BIC. This is done for a range of $\lambda$'s. The results are shown in figure \ref{fig:IC_error_vergleich}. While the BIC based on $L_{int}$ changes every time the corresponding number of parameters changes, the BIC based on $L_{ls}$ is monotonically increasing and only changing at some of the parameter number changes for this problem. The minimal plateau of the least squares error does not contain an interval where the correct number of model terms is found as a solution and the error even increases exactly when we reach the correct number of parameters showing that the norm based on the integrated model is superior in picking a sparse and model. Beside this the plateaus limit our toolbox for minimization since e.g. gradient-based methods will get stuck.  On the other hand, using an approach like Nelder-Mead, which is not gradient-based runs into similar problems. Therefore we are implementing a Bayesian optimization procedure to minimize the potential \cite{hyperopt}. The implementation \cite{data} of the new proposed algorithm  is in its simplest form explained in Table \ref{alg:bayesian_optimization}.

\begin{algorithm}
\caption{REBEL}\label{alg:bayesian_optimization}
\begin{algorithmic}[1]
\Require Time-series data, sampling rate, initial hyperparameters
\Ensure Sparse and accurate dynamical system model
\State Choose $\lambda$ to be one or multidimensional
\State Define ranges for $\lambda$ and $\alpha$
\State Preprocess data to estimate time derivatives
\State Initialise bayesian optimisation
\State Retrieve initial estimate for $\lambda$ and $\alpha$
\For{each iteration of Bayesian optimization}
    \State Estimate $\sigma_{est}$ via thresholded least squares with $\lambda$ 
    \State Integrate model based on $\sigma_{est}$
    \State Calculate BIC from $L_{mixed} = \alpha L_\text{int} + (1-\alpha)L_\text{W2}$
    \State Update hyperparameters based on BIC
\EndFor
\State Select model with minimal combined error
\State Return identified dynamical system model
\end{algorithmic}
\end{algorithm}

This algorithm depends on the crucial point of determining a range of values of $\lambda$ that should be considered by the Bayesian optimization. 
The problem of selecting a range for $\lambda$ can be explored by examining two extreme cases. If $\lambda$ is set lower than the smallest value from the first least squares fit, no terms are removed during thresholding. As a result, the iterations will always produce the same outcome, leaving the full library of terms intact, including spurious ones.
On the other hand, if $\lambda$ is set higher than the largest value from the first least squares fit, all terms are discarded immediately after the first fit. This leads to a model with no terms, and since all terms are removed in the first iteration, subsequent iterations will also produce an empty model.
These considerations also lead back to our critic of \cite{Mangan2017}, where the range of $\lambda$ excludes all models that are overfitting, i.e. more terms than in the original equation, and only focuses on a range of $\lambda$, where we get fewer terms and therefore obviously get wrong models.

Before demonstrating the capabilities of this algorithm in complex applications, we briefly address its performance under varying data sampling rates. To evaluate this, we compared the performance of the SINDy method with our newly developed method across different sampling ratios. Specifically, we estimated a model for the van der Pol equation at various sampling rates and then reintegrated the estimated models. As shown in figure \ref{fig:IC_error_vergleich}, our method consistently outperforms the SINDy method across all sampling rates in terms of the L2-norm of the difference between the integrated time series and the original data. This highlights its robustness in scenarios where fine-grained sampling is not achievable.

%
\section{Results}
To show the capabilities of the method we apply it to two problems from different fields of science. The first example deals with estimating a model for a low dimensional representation of the famous cylinder wake problem in fluid dynamics while in the second example the algorithm is used to estimate  models for latent spaces of fMRI data. In both cases we not only reduce the number of estimated coefficients significantly, but also improve the overall reconstruction error.

\subsection{A Low order model of a Cylinder Wake\label{sec:cylinderwake}}

Due to the nonlinear and non-local nature of the Navier-Stokes equation deriving low order models for certain flow configurations is difficult. A benchmark problem for low order modeling is the wake flow past a cylinder. In this setting the laminar fluid flow is distorted by a cylinder and then exhibits periodic vortex shedding behind the obstacle, which is a well-known phenomenon called von-K\'arm\'an vortex street \cite{Karman1911,Karman1912}. This system can be modeled by a Hopf bifurcation at the onset of the vortex shedding for a Reynolds number of about $Re\approx47$ \cite{Jackson1987,Zebib1987}. This system was a long standing problem for Galerkin-modelling, since the Hopf bifurcation could not be captured in a simple  POD-Galerkin-expansion. Since the Navier-Stokes equations are quadratic at least 3 POD modes are needed to capture a Hopf bifurcation, but it was shown that at least 8 POD modes were needed to find the correct physical behavior, i.e. a stable limit cycle. 
\begin{figure*}[ht]
\includegraphics[width=\linewidth]{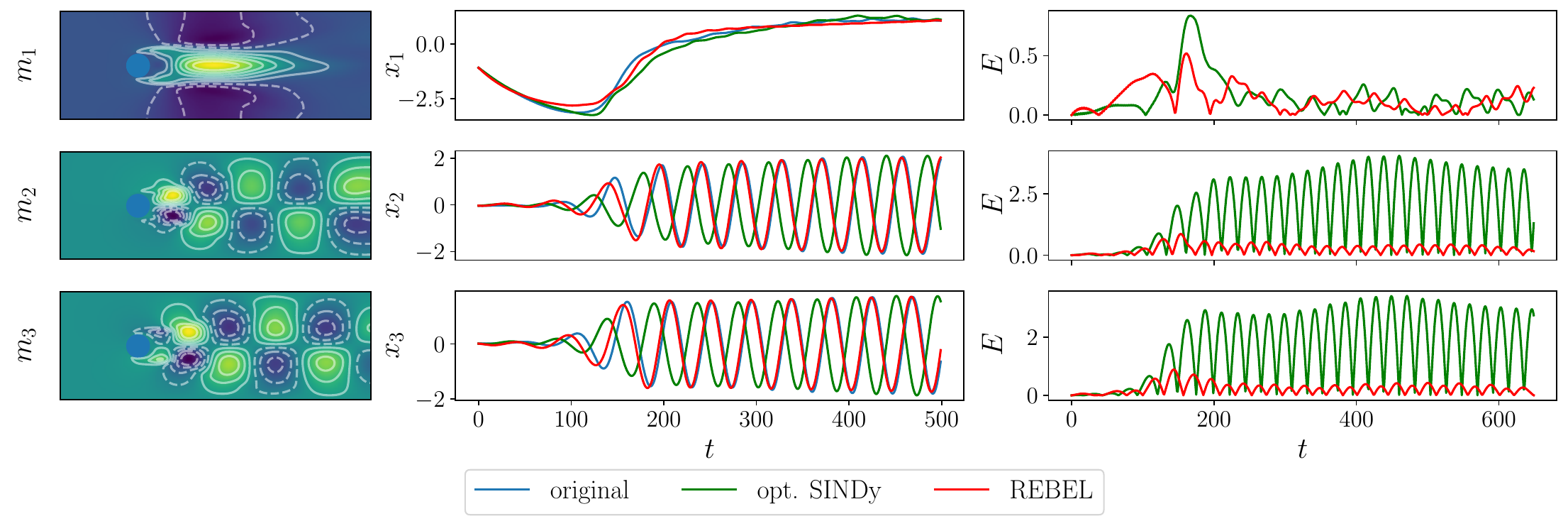}
\caption{Results for the analysis of the cylinder wake. (left) From top to bottom the first three POD modes of the problem are shown. (middle) Comparison of original data with the best estimated models from an optimized version of SINDy and REBEL for the three time series of the first POD modes. While the optimized SINDy model is able to capture a transient going into a limit cycle, the limit cycle has the wrong phase. The REBEL model captures the transient into a limit cycle and has the correct phase. (right) Comparison of of the difference between original data and estimated models. The error in the first mode shows that both optimized SINDy and REBEL can model the transient, but the optimized SINDy method has bigger peaks in the error. The errors over time for the second and third mode show that the error of the best REBEL model outperforms the error of the best optimized SINDy model significantly and the high peaks of the optimized SINDy error signalise the wrong phase, which is almost in antiphase to the original data.}
\label{fig:cylinder_vergleich}
\end{figure*}
A famous attempt to solve this problem is the semi-analytical 3-dimensional model of Noack et.al. \cite{Noack2003jfm}. Here a \textit{shift mode} was constructed, which, together with the first two POD modes, was sufficient to get a first physical Galerkin model of the vortex shedding. \cite{Brunton2016} showed that applying SINDy to the first two POD modes and the shift mode from \cite{Noack2003jfm}, estimates the Galerkin model from \cite{Noack2003jfm} for a certain choice of the sparsity parameter. While this attempt was able to capture a stable limit cycle, the phase of the model was incorrect since SINDy is agnostic to the realised temporal evolution of the system.
We performed a POD of a full simulation of the cylinder wake for a Reynolds number of $Re=100$ and projected the data on the first three modes. The resulting time series are similar to the three dimensional problem in \cite{Noack2003jfm}, since a POD on data including the transient contains the shift mode (cf.\cite{Noack2016}). The library $\Xi$ was set up to contain polynomials up to order three.

Using the before introduced method we estimate a model of the cylinder flow and compare it to the minimal BIC for the least squares error,  where the second model corresponds to the best SINDy model estimated with BIC as a metric denoted as "opt. Sindy" for "optimized SINDy". By this we mean that we use our bayesian hyperparameter optimization to find a value of the thresholding parameter $\lambda$, that corresponds to a minimal least squares error $L_{ls}$ based on the time derivatives, in order to utilise the optimization for the choice of $\lambda$ instead of choosing it by hand. In figure \ref{fig:cylinder_vergleich} the estimated and the original time series are shown.

\begin{figure*}[ht]
\includegraphics[width=\linewidth]{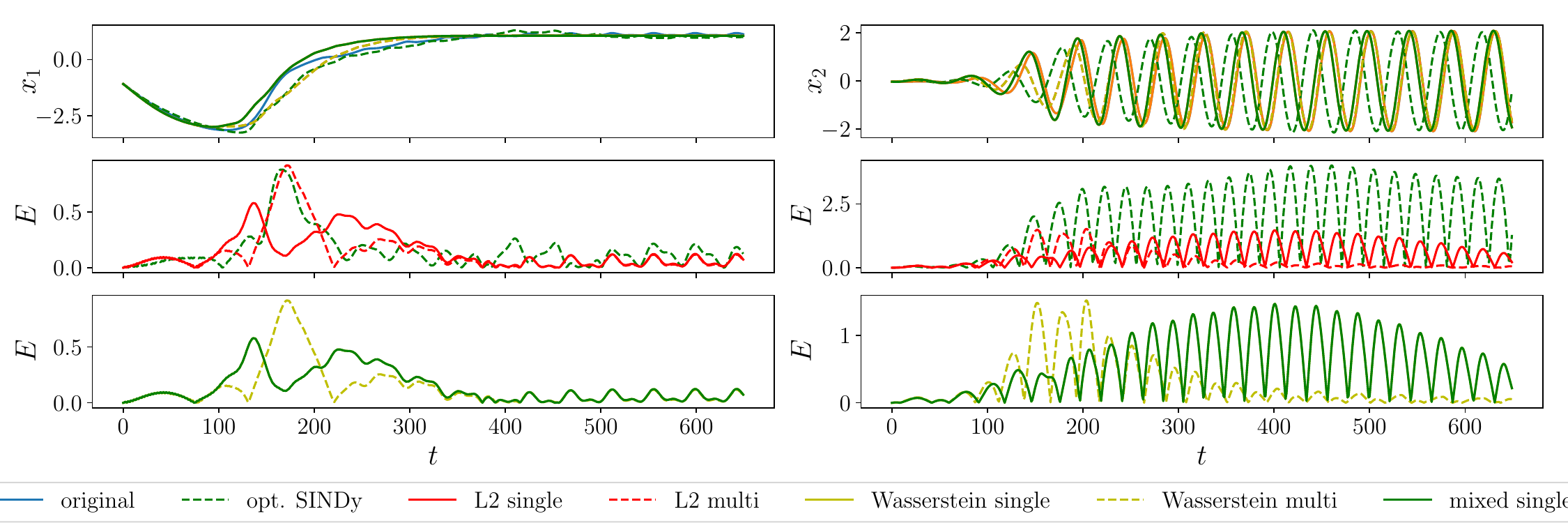}
\caption{Errors of the different estimated models compared to the original timeseries for different versions of BIC and used error functions. The choices for the models are the following:
(sindy) The best model was found for the $\lambda$ that minimizes the BIC of the least squares error (L2 single) The BIC of the L2 norm of the integrated timeseries was used with a single$\lambda$ (L2 multi) The same optimization as in L2 single was used but with multiple $\lambda$s  (Wasserstein single) The BIC of Wasserstein metric of the integrated timeseries was used with a single$\lambda$ (Wasserstein multi) The same optimization as in Wasserstein single was used but with multiple $\lambda$s (mixed single) To estimate a model we used a mix of the L2 norm and the Wasserstein metric of the integrated timeseries with a single $\lambda$  }
\label{fig:cylinderCompare}
\end{figure*}

As expected the reconstruction error based estimation gives a model closer to the data. Also the number of parameters is with 20 roughly half the size than the 45 parameter model based on the least squares error. It is important to notice that while the first approach correctly identifies the slow manifold of the Hopf bifurcation the timeseries are almost in opposite phase to each other. This incorrect phase was also found in other attempts to extract a dynamical system for the cylinder flow from data with the SINDy method \cite{Taira2020}. But especially when using data driven models to control a system the phase can be a crucial parameter, e.g. when controlling coupled oscillators in the power grid \cite{Korda2018}.\\
In order to investigate the advantages of all the improvements of our method, we also included error plots for all eight different model choices (SINDy, L2-norm, Wasserstein-metric, mixed norm and single or multiple $\lambda$) in figure \ref{fig:cylinderCompare}. It can be seen that every newly introduced idea improves on the overall error of the model. The estimated errors together with the models can be found in the supporting information.
In this problem we see one major advantage of our method when comparing with SINDy respectively with simply using the least-squares error. The temporal information in the least-squares error is only implicit, as we can shuffle all data points and get the same model, if we keep the left and right hand side consistent. Only when integrating the equation we recover information about the temporal evolution of the system. A similar problem estimating an optimal model based on the $L_\text{ls}$ involving limit cycles, where the phase was not recovered correctly, is found in \cite{Dam2017}.

\subsection{Low order models of experimental fMRI data}

\begin{figure*}[ht]
\begin{center}
\includegraphics[width=1.0\linewidth]{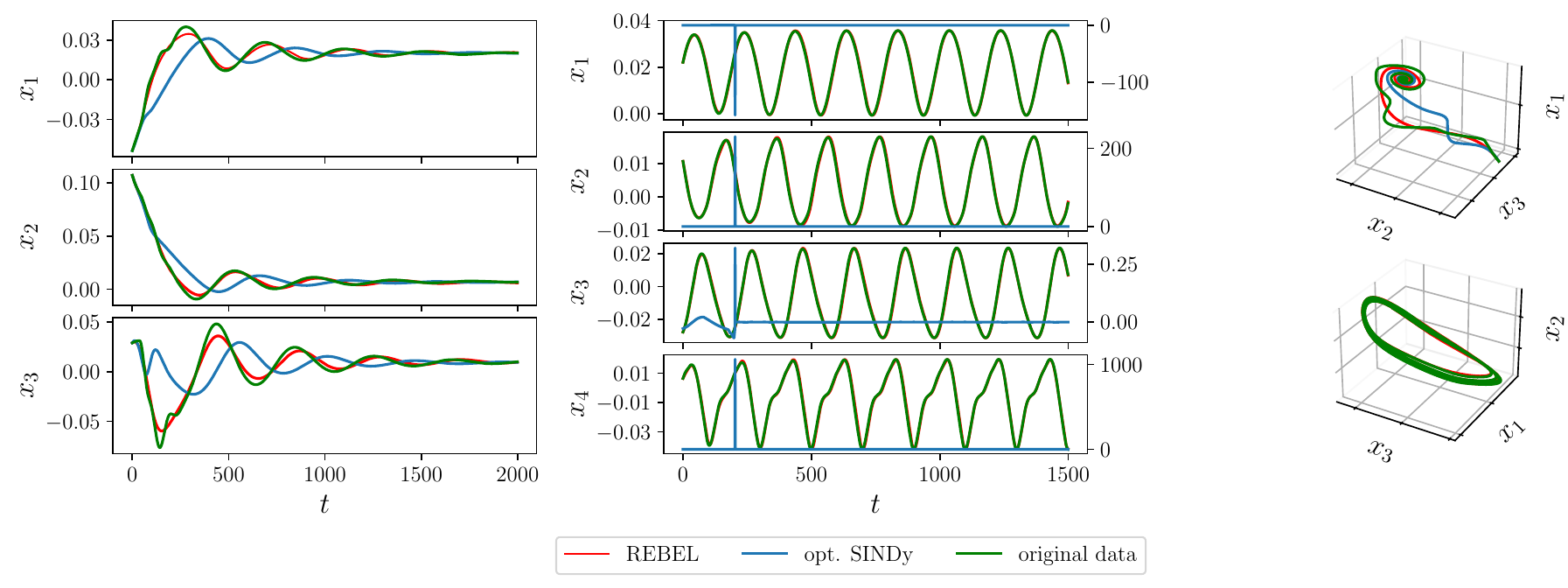}
\caption{(left) Comparison of integrated time series of best estimated models from the optimized SINDy and REBEL for the fixed point example of FMRI data. Both models do not perfectly match the original time series, but while the optimized SINDy method gives a different phase the REBEL method approximates most of the correctly. (middle) Comparison of integrated timeseries of best estimated models from optimized SINDy and REBEL for the limit cycle example of FMRI data. The optimized SINDy model is not able to estimate a correct model but diverges. For the optimized SINDy model we used a different scale on the right and used another scale for the REBEL method and the original data. The REBEL model is almost perfectly representing the data. (right) Comparison of the attractors of the estimated models for both datasets to the original data. For the fixed point example we see that REBEL approximates the data reasonably well, while the optimized SINDy method is going faster into the fixed point. For the limit cycle only the attractors of the original data and our REBEL model are shown and are in almost perfect agreement.}
\label{fig:fMRI}
\end{center}
\end{figure*}

The application of mathematical modeling in medicine has garnered significant attention over the past decades. For instance, in \cite{Davis2019}, health and disease are conceptualized as features of dynamical systems. Mathematical and data-driven methods have been applied to various medical fields, including Parkinson's disease and its treatment \cite{Tass2003,Adamchic2014}, cardiology \cite{Luther2011}, neuroscience \cite{Izhikevich2006}, and psychology \cite{Jamalabadi2022}, among others. In general, the use of mathematical models in medicine holds great promise for deepening our understanding of diseases and potentially improving treatments. For example, dynamical systems theory offers a robust mathematical framework for analyzing neurobiological processes \cite{Durstewitz2023nature}.  In \cite{Koppe2019}, for example, recurrent neural networks were employed to infer low-dimensional latent trajectories from high-dimensional experimental fMRI data obtained from human patients. Although the low dimensional trajectories generated by the neural representation show the typical solution structure of nonlinear ordinary differential equations, like limit cycles, the relaxation to fixed points or chaotic dynamics, the network is still a black-box giving no direct insight into the mathematical structure of the dynamics. In order to draw more insights from the estimated latent trajectories, we apply the REBEL-method to two examples of fMRI data, i.e. latent space trajectories from \cite{Koppe2019}. In general, estimating a dynamical system for the low-dimensional model adds a further reduction of complexity to the model not in terms of dimension but instead on the level of model parameters.

Following \cite{Koppe2019} a recurrent neural network was trained with a 6-dimensional latent space on a 20 dimensional data set and then it was used to generate new timeseries in the low dimensional latent space. 
For the first patient fixed point dynamics were found. The number of timeseries was again reduced with a proper orthogonal decomposition (POD) to the first three POD modes. As before a sparse model was estimated based on the BIC for both error functions. The ansatz for the dynamics was chosen as  polynomials up to order two in all three variables. The reconstruction of the two models in comparison to the original data is shown in figure \ref{fig:fMRI}. While both models run into a fixed point, only the reconstruction error based model is able to capture the first part of the relaxation process while needing again fewer model terms.

For the second patient the trained network exhibits a higher dimensional limit cycle. The data was projected onto the first four POD modes. Again an ansatz with terms up to order three was used. The reconstruction of the two models in comparison to the original data is shown in figure \ref{fig:fMRI}. In this case the least squares based estimation indeed gives a sparse model, but the forward calculation of the model diverges at some point. On the other hand the reconstruction error based model gives a good reconstruction of the limit cycle over a considerable temporal region with fewer model terms. 

The recurrent neural network here is defined as a difference equation for the variables in the latent space (cf. page 4 in \cite{Koppe2019}). Since this network works on subsequent data points to generate the new latent space trajectories, we basically estimate the right hand side of a difference equation with our method. As our estimation is resulting in a symbolic expression of the neural network dynamics, this can be linked to efforts in the field of explainable AI like SHAP-Values \cite{Lundberg2017}. In fact, through the estimation of a sparse differential equation for the output of a recurrent neural network, we can distinguish between relevant and irrelevant inputs, which is also one of the main goals of SHAP-Values.

Following Wiener \cite{Wiener1961} we could see this as a kind of gray-box model, since we combine the black-box model, i.e. the RNN, with a white-box model, i.e. our differential equation. Additionally, we can look at the estimation of a dynamical system for the low-dimensional RNN-model as further reducing complexity. This reduction is not in terms of dimension but instead on the level of model parameters, leading potentially to better explainability.
\\ 
In this special example of a limit cycle in fMRI data, this has a direct importance. Since the RNN relies on a finite step size, even a stable limit cycle could seem unstable in the long term. As our method in contrast directly estimates the right hand side of the neural network, the stability is directly linked to the resulting differential equation and can be calculated without the problem of finite step sizes. This addresses a problem, which is also recognized in \cite{Durstewitz2023nature}. Often data-inferred RNNs can be generative in the sense, that when sampled out of data, we recover similar statistics for e.g. relevant physiological or neurological data, but we cannot be sure that the long term behavior of the RNNs are correct for out of sample data.\\

\section*{Discussion}
In this paper we introduce a novel method to estimate differential equations for dynamical systems in a way, that exceeds previous approaches in terms of accuracy and sparsity. The latter directly linking our approach to better explainability.

Starting with the observation that the quality of the time derivative, the available sampling rate, and their interrelation pose significant challenges to both the sparsity and accuracy of identifying dynamical systems from data, we develop a new method for identification of dynamical systems from data overcoming these issues.

Here, we address this problem by combining system identification based on a thresholded least squares minimization, as a convex optimization problem, with  new error measures. For the new measures we not only look at time-derivatives but explicitly take into account how the models perform when used for forecasting the dynamics, i.e. to integrate the model again in time. From these integrated time series we calculate the integrated least squares error \( L_{\text{int}} \) and the Wasserstein metric \( L_w \). By incorporating this approach into a Bayesian optimization routine, we efficiently determine the optimal hyperparameters for thresholding of the convex problem and the mixture of \( L_{\text{int}} \) and \( L_w \). This mixture is optimized to better tune the advantages and disadvantages of the two errors to the problem at hand. Consequently, introducing a distinct regularization parameter for each individual differential equation of the in general n-dimensional system further improves the capabilities of our approach.

With this method, we were able to identify a very sparse and highly accurate model for the benchmark problem of the wake flow behind a cylinder, without using any a priori information. This model accurately captures the transient dynamics and the oscillation period of the generated vortices while reproducing the correct phase, which is another improvement compared to previous approaches.

In the example of fMRI data, we have shown how to gather more insight from trained neural networks. On the one hand our approach here has shown to work in an example, where the state of the art methods have problems. On the other hand we could demonstrate that for recurrent neural networks estimating a differential equation of their learned dynamics is an alternative to methods in the field of explainable AI. As we not only determine correlational relationships between inputs and outputs, but estimate differential equations, we are also able to capture possibly causal relationships. \\
In general we think that this approach has a wide range of applicability in the field of explainable AI to postprocess trained networks in order to gather more insights into the dynamics of different systems, especially but not limited to AI applied to dynamical systems.

Since our method relies on an information criterion based on the reconstruction error from the estimated models, it might work without fewer or even no a priori information about physical constraints. As we can in general not even derive those constraints for some systems, this should lead to better models.For our examples constraints like energy conservation are directly manifested in the temporal evolution of the time series and should therefore indirectly regularize the choice of models, whilst not directly imposing or even knowing them.




\bibliography{library.bib} 

\end{document}